\documentclass[prl,twocolumn,showpacs,superscriptaddress,10pt,longbibliography]{revtex4-1}

\usepackage {graphicx}
\usepackage {amssymb}
\usepackage {amsmath}
\usepackage {mathrsfs}
\usepackage{xcolor}

\newcommand{\InvisibleRemark}[1]{}

\newcommand{\InvisibleTodo}[1]{}
 \def\1#1{{\color{blue}#1}}

\newcommand{\reffig}[1]{Fig.~\ref{#1}}
\newcommand{\refeq}[1]{Eq.~(\ref{#1})}
\newcommand{\refeqs}[2]{Eqs.~(\ref{#1})-(\ref{#2})}

\newcommand{\pderiv}[2]{\frac{\partial#1}{\partial#2}} 
\DeclareMathOperator{\real}{Re}

\DeclareMathOperator{\tr}{tr}
\DeclareMathOperator{\sech}{sech}

\newcommand{\ket}[1]{\ensuremath{\left|#1\right\rangle}}

\begin{document}

\title{Waveshape tolerant photonic quantum gates}
    
\author{I. Babushkin}
 \affiliation{Institute of  Quantum Optics, Leibniz University Hannover,
   Welfengarten 1, 30167 Hannover, Germany}
\affiliation{Max Born Institute, Max-Born-Str.~2a, 12489 Berlin}
\affiliation{Cluster of Excellence PhoenixD (Photonics, Optics, and
  Engineering - Innovation Across Disciplines), Welfengarten 1, 30167
  Hannover, Germany}

\author{A. Demircan}
 \affiliation{Institute of  Quantum Optics, Leibniz University Hannover,
   Welfengarten 1, 30167 Hannover, Germany}
\affiliation{Cluster of Excellence PhoenixD (Photonics, Optics, and
  Engineering - Innovation Across Disciplines), Welfengarten 1, 30167
  Hannover, Germany}

\author{M. Kues}
\affiliation{Cluster of Excellence PhoenixD (Photonics, Optics, and
  Engineering - Innovation Across Disciplines), Welfengarten 1, 30167
  Hannover, Germany}
\affiliation{Institute of Photonics, Leibniz University Hannover, Nienburgerstr. 17, 30519 Hannover}

 \author{U. Morgner}
 \affiliation{Institute of Quantum Optics, Leibniz University Hannover,
   Welfengarten 1, 30167 Hannover, Germany}
\affiliation{Cluster of Excellence PhoenixD (Photonics, Optics, and
  Engineering - Innovation Across Disciplines), Welfengarten 1, 30167
  Hannover, Germany}

\date{\today}

\begin{abstract} 
  Photons, acting as ``flying qubits'' in propagation geometries such
  as waveguides, appear unavoidably in the form of wavepackets
  (pulses). The actual shape of the photonic wavepacket, as well as possible
  temporal/spectral correlations between the photons, play a critical
  role in successful scalable computation. Currently, unentangled
  indistinguishable photons are considered as a suitable resource for
  scalable photonic circuits. Here we show that using so called
  coherent photon conversion, it is possible to construct flying-qubit
  gates, which are not only insensitive to waveshapes of the photons
  and temporal/spectral correlations between them, but which also
  fully preserve these waveshapes and correlations upon the
  processing. This allows using photons with correlations and purity
  in a very broad range for a scalable computation. Moreover, such
  gates can process entangled photonic wavepackets even more
  effectively than unentangled ones.
\end{abstract}

\maketitle

Despite photons being seemingly ideal candidates for carrying quantum
information, robust scalable gates and circuits for the photonic
qubits remain yet illusive. The difficulty is that photons do not
interact to each other directly, and thus either indirect interaction
via a nonlinear medium
\cite{langford11,xia16,niu18,viswanathan15,viswanathan18,kounalakis18-cross-kerr-superconduct,tiarks19-rydberg,sagona-stophel20,heuck20,dot14,meyer-scott15,solntsev21}
or ``emulation'' of interaction using measurements
\cite{knill01,obrien03,kok07-rev,carolan15} are needed. The waveshapes
either in time
\cite{marcikic02-time-bins,xiong15-time-bin-exp,brecht15,ansari18-temporal-modes,luo19-electroopt-circuits}
or in frequency
\cite{ramelow09-freq-bins,olislager10-frequency-bins,reimer16} (or
both together \cite{kues17}) may themselves represent qubits.
Alternatively, information can be stored in degrees of freedom
orthogonal to waveshapes, such as which-path \cite{kok07-rev},
polarization \cite{kok07-rev}, field quadratures
\cite{braunstein05,andersen10-rev-cont-vars} or angular momentum
\cite{mair01-entanglement-oam}. In all cases, regardless of the used
approach, the shape of the photonic wavepacket is of primary
importance for the gate action.

One ideally would desire a gate, working with (almost) arbitrary
photonic waveshapes and fully preserve them during the computation --
even if the waveshapes are not unentangled or not pure. We will call
such gates ``waveshape tolerant''.
Formally speaking, waveshape tolerantness is a quite strong property;
For instance, if we delay one of the two photons in a two-photon gate
so that they do not overlap in time anymore, the waveshape tolerant gate
should still work. Are such gates possible at all?

In linear optical circuits (LOC)
\cite{knill01,obrien03,kok07-rev,carolan15} which consist of linear
interferometric networks and measurements of ancillary qubits, it is
known \cite{uren03,uren05} that for scalable operation the photons in
different channels have to be independent and indistinguishable, so
that no which-path information can be extracted, since its presence
breaks the interference in linear optical elements, and introduces
incoherence during measurements \cite{uren03,uren05,babushkin20}.
Although the action of LOC was recently extended to more general class
of states \cite{babushkin20}, they certainly are not waveshape tolerant in
the sense mentioned above.

The situation seems even worse if we consider a recent proposal based
on so called coherent photon conversion (CPC), where the photon-photon
interaction is based on a four wave mixing (FWM) process in presence
of a strong coherent laser field \cite{langford11}. This coherent
field amplifies the action of the $\chi^{(3)}$ nonlinearity between
the remaining three waves which are in a Fock state. A two-photon
wavepacket sent to the input of a CPC device should experience
up-/down-conversion cycles, attaining additional phase shift of $\pi$
after each cycle, thus leading to a nonlinear sign (NS) gate
functionality.
Unfortunately, soon after the discovery of this approach it was noted
\cite{viswanathan15} that if a finite size of the wavepackets is taken
into account, it does not work even if the participating photons are
temporally unentangled and indistinguishable, since only the parts of
the bi-photon wavepacket located at the same position in space (or
time) can effectively interact. This problem can be however solved if
the photons have nonzero group velocity relative to each other
\cite{xia16,viswanathan18}. In this case, parts of the photons which
were initially separated, eventually cross, and can interact at that
moment. This approach was very recently partially realized
experimentally \cite{sagona-stophel20} in a system with
electromagnetic-induced transparency, however at the cost of partial
loss of ``working'' photons. Besides, three photon interaction was
observed in cold Rydberg atoms \cite{liang18-3photon}.

Nevertheless, with or without a relative movement of the photonic
wavepackets, it seems at the first sight, that the CPC-based approach
has no chances to produce waveshape tolerant gates, since, by up-conversion
of a bi-photon wavepacket (described by a two-dimensional
distribution) into a single photon (described by a one-dimensional
waveform), some information about the initial two-photon waveshape
is unavoidably lost, and the back-conversion cannot restore it
anymore. Therefore, the final waveshape seemingly can coincide with
the initial one only by chance.

\begin{figure}[t!]
  \includegraphics[width=0.5\textwidth]{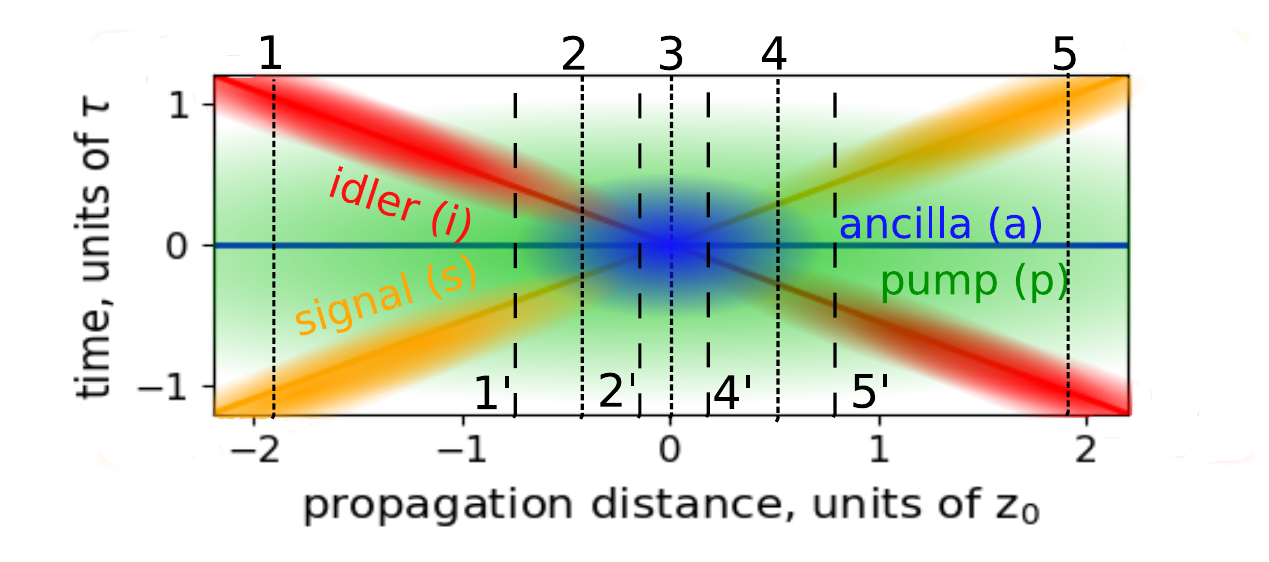}
  \caption{
    CPC-based nonlinear sign (NS) gate with group-velocity-unmatched pulses.
    Single photon wavepackets denoted as ancilla ($a$, blue), signal ($s$, orange)
    and idler  ($i$, red) propagate in presence of strong coherent pump 
    ($p$, green), interacting via the FWM process. The moving
    centers of the wavepackets are indicated by solid lines. Color
    change denotes schematically conversion of photons. 
    Vertical lines and numbers correspond to the positions of the
    snapshots shown in \reffig{fig:dyn} (lines labeled 1-5 refer to 
    S1,S2,S4, lines labeled 1',2',3,4',5' refer to S3). }
\label{fig:base}
\end{figure}

Despite of these objections, here we show that CPC-based gates can
indeed posses the
property of being waveshape tolerant, if the wavepacket is slowly
varying on the time scale of the effective photon-photon interaction,
or, equivalently, if the suitable fourth-order coherence time is large
enough. This implies also scalability of the gates to many entangled
photons, as well as their ability to work with mixed states. Moreover,
we show that processing of time- and frequency-entangled photons can
be more efficient than of the unentangled ones.

\begin{figure*}[t!] 
\includegraphics[width=1.0\textwidth]{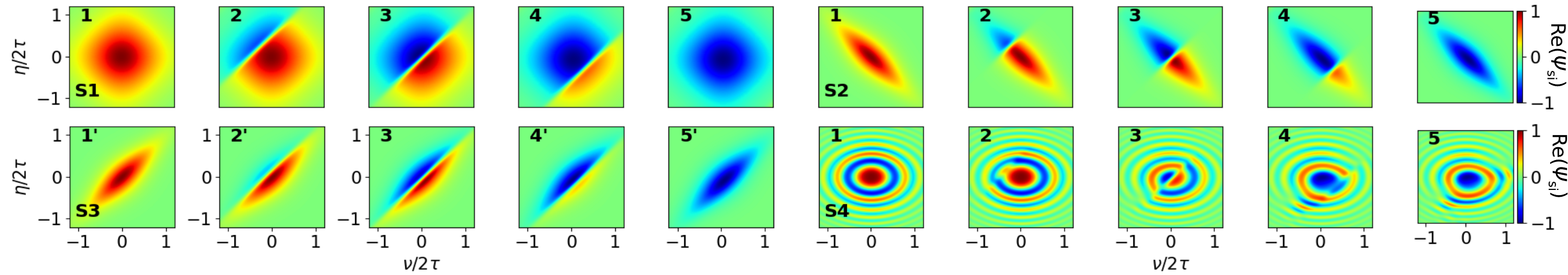} 
\caption{Snapshots of the signal idler wavepacket represented by
  $\real(\Psi_{\mathrm{si}}(t_s,t_i,z))$ for different initial
  waveshapes marked as S1...S4 (see text) shown in the moving
  coordinates $\eta=(t_s-z\beta_{1s})$, $\nu=(t_i-z\beta_{1i})$. The
  positions of the snapshots in $z$ are indicated in \reffig{fig:base}
  by the vertical lines and corresponding numbers. Sign flip of
  $\real(\Psi_{\mathrm{si}})$ 
  indicates successful NS gate operation. The waveshapes in S1...S3
  are fully conserved, which is not the case for S4 where distortions
  appear.}
\label{fig:dyn}
\end{figure*}

\textit{The model.} The CPC approach with group velocity-unmatched pulses
proposed in \cite{xia16} is shown in \reffig{fig:base}. 
The photons in Fock states at 
frequencies $\omega_s$, $\omega_i$, $\omega_a$, which we
call signal ($s$), idler ($i$), and ancilla ($a$), respectively, the first two (inspired by $\chi^{(2)}$
terminology \cite{boyd92:book}), interact via the FWM process described
by the susceptibility
$\chi^{(3)}=\chi^{(3)}(\omega_a;\omega_p,\omega_s,\omega_i)$ in presence
of a strong coherent field ($p$) at frequency $\omega_p$ with the peak
electric field $E_p$. This induces an effective three wave mixing (TWM)
interaction between the $a$, $s$ and $i$ wavepackets. 
Each of the wavepackets
is described by the frequency-dependent operators $a_a(\omega_a)$,
$a_s(\omega_s)$ and $a_i(\omega_i)$. 

Since only parts of the wavepackets which are close to each other (in
space or time) can interact, one has to resolve the finite size of the
interaction region \cite{xia16}. In the previous considerations
\cite{xia16,viswanathan18}, the Hamiltonian containing spatial
nonlocality was used for this purpose, attracting however critics as
physically nontransparent \cite{viswanathan18}. Here we operate
instead in terms of non-instantaneous response, that is, with
nonlocality in time rather than in space, which is more physically
justified (see \cite{bloembergen96:book} and also below), leading
nevertheless to a formulation rather equivalent to
\cite{xia16,viswanathan18} in our particular case. 
The non-instantaneous interaction can be represented with a Hamiltonian
\begin{gather}
  H = \hbar \gamma'   \int \tilde h(\omega_a,\omega_s,\omega_i)a_s^\dag(\omega_s) a_i^\dag(\omega_i)
  a_a(\omega_a) \times \nonumber \\
  \times d\omega_sd\omega_id\omega_a  + \mathrm{h.c.}.
  \label{eq:h_omg}
\end{gather}
Here $\hbar$ is the Plank constant, $\gamma'\sim \chi^{(3)}E_p$ is the
interaction strength (see more details below, after
\refeqs{eq:main-eqs-p}{eq:main-eqs-si}), and $\tilde h$ describes the
non-instantaneous response of the nonlinear medium in frequency
domain. Although our further analysis shows that the particular form
of $\tilde h$ is not important, for numerical simulations we use the
function
\begin{equation}
\tilde h(\omega_a,\omega_s,\omega_i) = \frac{\delta(\omega_p-\omega_a+\omega_s+\omega_i)}{2\pi}
e^{-\sigma^2(\omega_s^2 + \omega_i^2)/2},\label{eq:htilde}
\end{equation}
where $\delta$ is the Dirac $\delta$-function. This response function
has a transparent physical meaning: $\delta$-function ensures the
energy conservation; The frequency-dependent form-factor effectively
``switches off'' the interaction at frequencies larger than
$\omega_R\approx 2\pi/\sigma$. This represents schematically a typical
frequency-dependent $\chi^{(3)}$ for dielectrics
\cite{boyd92:book,bloembergen96:book,bree12:book}, which vanishes at
frequencies above the band-gap. The presence of the cutoff frequency
$\omega_R$ in \refeq{eq:htilde} means that the photons can effectively
interact to each other only if they the are separated in time by an
interval
smaller than $\sigma$. This implies also spatial nonlocality:
The photons interact if they are at the distance smaller than
$c\sigma$. For ``typical'' materials like fused silica $\sigma$ has
been recently measured to be at the level of hundreds of attoseconds
\cite{hofmann15,sommer16}.

The equation of motion can be written in the form of the Schr\"odinger
equation $i\hbar\partial_t\ket \Psi=(H_0+H)\ket \Psi$, where $H_0$ is
the non-interacting Hamiltonian. It is useful to formulate the
corresponding wavevector $\ket \Psi$ in terms of temporal modes 
\cite{brecht15}:
\begin{gather}
\nonumber \ket \Psi =
\int\int\Psi_{\mathrm{si}}(t_s,t_i,z)\ket{t_s}_s\ket{t_i}_idt_sdt_i
+ \\ + \int\Psi_a(t_a,z)\ket{t_a}_adt_a,
  \label{eq:psi_cont}
\end{gather} where
$\ket{t_j}_j=\int e^{it_j\omega_j}a_j^\dagger(\omega_j)
\ket{0}_{\omega_j} d\omega_j$, $j=\{s,i\}$,
$\ket{t_a}_a=\int e^{it_a
  (\omega_a-\omega_p)}a_a^\dagger(\omega_a)\ket{0}_{\omega_a}
d\omega_a$, where $a^\dagger$ and $\ket{0}$ denote creation operators
and vacuum states for the corresponding modes and frequencies. Also,
we assume that the higher order linear dispersion is negligible
leading to
$H_0=\sum_j \hbar \int \omega a_j^\dag(\omega)a_j(\omega)d\omega + \mathrm{h.c.}$ (see
Supplementary).

This allows us to write down the evolution equations for
$\Psi_{\mathrm{si}}(t_s,t_i,z)$, $\Psi_a(t_a,z)$ in $z$-direction in space,
assuming 1D propagation geometry such as for example in waveguides.
$t$- and $z$- formulations are mathematically equivalent in our case
\cite{quesada20-quant-prop-eqs} (no higher order dispersion and small
group mismatch, see Supplementary for more details). The resulting
equations written in the frame of reference propagating with the group
velocity of the ancilla $v_a$ are:
\begin{widetext}
  \begin{gather}
    \label{eq:main-eqs-si}
    \pderiv{\Psi_{\mathrm{si}}(t_s,t_i,z)}{z} =
    -\beta_{1s}\pderiv{\Psi_{\mathrm{si}}(t_s,t_i,z)}{t_s}
    -\beta_{1i}\pderiv{\Psi_{\mathrm{si}}(t_s,t_i,z)}{t_i} - i\gamma  \int
    h(t_a,t_s,t_i)\Psi_a(t_a,z)dt_a, \\
    \pderiv{\Psi_a(t_a,z)}{z} =  -
    i\gamma \int\int h(t_a,t_s,t_i)\Psi_{\mathrm{si}}(t_s,t_i,z)dt_sdt_i.
    \label{eq:main-eqs-p}
  \end{gather}
\end{widetext}
Here $\beta_{1j}=1/v_j-1/v_a$, $j=\{s,i\}$, $v_j$ are the
corresponding group velocities, $h(t_a,t_s,t_i)$ is the Fourier
transform of $\tilde h(\omega_a,\omega_s,\omega_i)$:
\begin{equation}
  h(t_a,t_s,t_i)=\frac{1}{2\pi\sigma^2}e^{-(t_s-t_a)^2/2\sigma^2}e^{-(t_i
  -t_a)^2/2\sigma^2}.\label{eq:h}
\end{equation}

$\gamma\approx\gamma'/\sqrt{v_a}$ can be obtained from the limit of
instantaneous nonlinearity \cite{drummond14:book}  (see details in
Supplementary) as $\gamma = \hbar \omega^2_pn_2\Phi_p/cS$, where $S$
is the effective area of the beam (assumed to be the same for all
waves), $c$ is the speed of light in vacuum,
$\Phi_p=\sqrt{I_p/\hbar\omega_p}$, $I_p$ is the intensity of $p$, and
$n_2$ corresponds to $\chi^{(3)}$ of the relevant FWM process.

\textit{Numerical simulations.} In numerical simulations we assumed $\beta_{1s}=-\beta_{1i}\equiv \beta_1$, that is, the signal
and idler pulses propagate with equal velocities of opposite sign in
the frame of reference of the ancilla (see \reffig{fig:base}). If we
denote the initial pulse duration by $\tau$, the effective overlap (and thus
interaction) of the signal and idler waveshapes takes place over the
distance $\approx z_0$, where $z_0=\tau/\beta_1$.  For the sake of generality, 
we normalized time $t$ to $\tau$ and $z$ to $z_0$, leading to the 
following renormalizations in \refeqs{eq:main-eqs-si}{eq:main-eqs-p}: $\beta_{1j}\to 1$ for $j=\{s,i\}$, 
$\hbar\gamma\to\hbar\gamma /\beta_1$, 
$\Psi_a\to \Psi_a/\sqrt{z_0}$, $\Psi_{\mathrm{si}}\to \Psi_{\mathrm{si}}/z_0$.
Numerical simulations were made  in the range $z=[-L/2,L/2]$ for
$L=4.4z_0$
(see
\reffig{fig:base}). We initialized our waveshapes at $z=-L/2$ with
the vacuum for the ancilla ($\Psi_a=0$) and two photons in signal and idler
modes with various initial distributions
$\Psi_{\mathrm{si}}(t_s,t_i,-L/2)=\Psi^{\mathrm{(in)}}(t_s,t_i)$
for four different simulations denoted as S1$\ldots$S4 in
\reffig{fig:dyn}. For the simulations S1, S2, S3 we used the waveshape
\begin{equation}
  \nonumber
 \Psi^{\mathrm{(in)}}(t_s,t_i) = \Psi_0 \mathcal R_\phi\left[\sech{\left(\frac{t_s-t_{s0}}{\tau_s}\right)}\sech{\left(\frac{t_i-t_{i0}}{\tau_i}\right)}\right],
\end{equation}
where $\Psi_0$ is a normalization factor leading to
$|| \Psi_{\mathrm{si}}(t_s,t_i,-L/2)||=1$ (here $||f||$ is the norm of
a function $f$: $||f||=\int\int f(x,y) dx dy$), $\mathcal R_\phi$ is
the transformation rotating $\Psi(t_s,t_i)$ in $(t_s,t_i)$ plane by
the angle $\phi$. For S1 we used (in normalized units mentioned above)
$\tau_s=\tau_i=1$, $\phi=0$ which gives the separable waveshape, and
$t_{s0}=-1.2\tau_s$, $t_{s0}=1.2\tau_s$. For
S2 and S3 we assumed $\tau_s=1$, $\tau_i=1/3$ (in normalized units);
$\phi=\pi/4$ for S2 and $\phi=-\pi/4$ for S3, constructing thereby two
states with signal and idler being entangled. Finally, in S4 we
assumed the initial distribution as in S1 but modified by multiplying
$\Psi(t_s,t_i,-L/2)$ by the time dependent factor
$\exp{(iC_st_s^2+iC_it_i^2)}$ with $C_s=C_i/2=10$, introducing a
time-dependent chirp into both signal and idler pulses. In
simulations, we used $\sigma/\tau=0.05$.
As a ``reference case'' we take a fused silica waveguide with
$n_2= 3\times 10^{-16}$ cm$^2$/W, $S=1$ $\mu$m$^2$. We also
assume 
$\sigma$ of 500 attoseconds (as), which is of the order of values
measured recently \cite{hofmann15,sommer16}. The normalized parameters
mentioned above correspond then to $I_p=20$ TW/cm$^2$ and FWHM
duration of 10 fs ($\tau=5.4$ fs). The interaction distance is then on
the kilometer range: taking 
$z_0=3$ km ($\beta_1=1.8$ fs/km) leads to the fidelity
$\mathcal F=99\%$. On the other hand, faster than optimal propagation
reduces back-conversion and thus fidelity; for instance, increasing
$\beta_1$ two times 
leads to
$\mathcal F=90\%$. We note
furthermore that the gate length is inversely proportional to $n_2/S$;
Using of novel photonic materials
\cite{agrawal:book,leuthold10,motojima19,trojanek10,michinobu05,esembeson08},
such as silicon, noncrystalline diamond and others may reduce the gate
length to a centimeter range. For further discussion of these issues
see Supplementary.

The results of simulations are shown in \reffig{fig:dyn} as
snapshots of  
$\real \Psi_{\mathrm{si}}(t_s,t_i,z)$ at the positions marked by 1...5 (for S1...S3) and
1',2',3, 4', 5' 
(for S3) in \reffig{fig:base} by vertical lines. A successful
gate operation assumes the phase change of $\pi$, which means sign flip of
$\real \Psi_{\mathrm{si}}(t_s,t_i,z)$. As one can see in \reffig{fig:dyn}, the gate
operation takes place in a form of a ``front'' of the size $\sim\sigma$ which
moves through the waveshape.  In the case S3 the interaction is completed at
around 3 times shorter distance corresponding to the 3 times smaller duration
of the wavepacket in the direction of the front movement. 
Interestingly, at the point ``3'' we have
a Bell-type state $\ket{+}_s\ket{+}_i-  \ket{-}_s\ket{-}_i$, where $\ket{\pm}_{j}$ 
denotes the signal ($j=s$) or idler ($j=i$) photon,  
located in upper-left ($+$) or lower-right ($-$) corner of the $(\nu,\eta)$ plain. 

\begin{figure}[t!]
\includegraphics[width=0.5\textwidth]{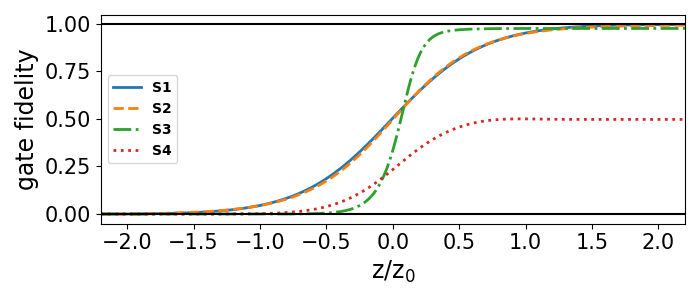} 
\caption{Evolution of the gate fidelity $\mathcal F(z)$ for the
  simulations S1...S4 in \reffig{fig:dyn}. }
\label{fig:fidelity}
\end{figure}

The quality of the gate operation and its ability to
keep the waveshape can be quantified by the fidelity $\mathcal F$
presented in \reffig{fig:fidelity} and defined as
$\mathcal F(z)=\dfrac12
\left|1-\int\int
  \Psi^{(\mathrm{norm})}_{\mathrm{si}}(t_s,t_i,z)\Psi^{\mathrm{(in)}*}(t_s,t_i)
  dt_sdt_i\right|$,
where
$\Psi^{\mathrm{(norm)}}_{\mathrm{si}}(t_s,t_i,z) =
\Psi_{\mathrm{si}}(t_s,t_i,z)/||\Psi_{\mathrm{si}}(t_s,t_i,z)||$. $\mathcal F=1$
corresponds to a perfect gate operation, including full conservation
of the pulse shape. One can see indeed from \reffig{fig:dyn} and
\reffig{fig:fidelity}, that for the cases S1, S2, S3 not only the
phase is successfully flipped, but the waveshape also remains intact
after the gate. On the other hand, the phase deformation introduced in
(S4) makes the waveshape changing quickly on the scale of $\sigma$,
leading to a visibly corrupted waveshape reflected in the low fidelity
in \reffig{fig:fidelity}.

Summarizing, smooth enough wavepackets are processed in a
waveshape tolerant way. On
the other side, processing of waveshapes with features, ``sharp'' on
the scale of $\sigma$, is not anymore perfect.

\textit{Analytics.} We proceed further with a more general
analytical
insight. We derive the equation describing evolution of the
wavepacket $\Psi_{\mathrm{si}}$ for an arbitrary response function $h$
localized mostly in the region of the size $\approx\sigma\times\sigma$
of the point $t_s=t_i$ and being symmetric in respect to its maxima
($h$ used above is a particular example of such function), whereas
$\Psi_{\mathrm{si}}$ is slow on the scale of $\sigma$, meaning:
\begin{equation}
  |\partial_{t_j} \Psi_{\mathrm{si}}(t_s,t_i,z)|\ll |\Psi_{\mathrm{si}}(t_s,t_i,z)|/\sigma
  \label{eq:slow}
\end{equation}
for $j=\{s,i\}$ for all $t_s$, $t_i$. 
Under these conditions, the equation describing the evolution of a small part
$\delta \Psi$ of the size $\sigma\times \sigma$ of the whole
wavepacket $\Psi_{\mathrm{si}}$ is described by (see Supplementary for more
details):
\begin{equation}
  \label{eq:internal-eq}
  \partial_{\xi\xi} \delta\Psi(\xi,\eta,\nu) + \frac{\gamma^2}{\sigma} K(\xi-\xi_0) \delta\Psi(\xi,\eta,\nu) = 0,
\end{equation}
where $\xi=\beta_1(z-t_i/\beta_{1i} + t_s/\beta_{1s})$, $\eta=(t_s - z \beta_{1s})$,
$\nu= (t_i-z\beta_{1i})$,  $K(\xi-\xi_0)$ 
is a coupling factor which is of the order of one 
close to $\xi=\xi_0$ and quickly approaches zero as $|\xi-\xi_0|$ becomes larger than  
$\sigma$ (that is, we leave the $\sigma$ vicinity of $\xi_0$).
For instance, for a $h$ given by \refeq{eq:h} we have
$K(\xi-\xi_0) \propto  \exp{\left(-(\xi-\xi_0)^2/4\sigma^2\right)}$.
 
\refeq{eq:internal-eq} is 
an equation for a harmonic oscillator with variable frequency. In the
interaction region of the size $\sigma$ near $\xi=\xi_0$ the harmonic
oscillator (and thus the conversion) is effectively switched on, and
outside, it is off. This region defines the conversion front visible
in \reffig{fig:dyn}. For the perfect gate operation, exactly half of
an oscillation period is necessary.
Since \refeq{eq:internal-eq} is linear, and the evolution starts from
the same initial phase ($\Psi_{\mathrm{si}}$ is at maximum) in all points
$(\eta,\nu)$, the same part of oscillation is performed everywhere,
making the shape of $\Psi_{\mathrm{si}}$ preserved.

Importantly, this result is general in the sense that it applies for
any $\Psi_{\mathrm{si}}$ satisfying the slowness condition
\refeq{eq:slow}, also for arbitrary noninstantaneous response $h$
satisfying the conditions mentioned above. S4 in \reffig{fig:dyn}
shows that if this condition is not satisfied, the gate may indeed
modify the wavepacket shape, since the noninstantaneous interaction
mix different pieces of the waveshape.

\textit{Scalability and operation for mixed states.} Here we show,
that the NS gate works properly for mixed states (described by a
density matrix $\rho_\mathrm{si}$),  if a more general slowness
condition is satisfied. We sketch the consideration, technical details
can be found in the Supplementary. $\rho_\mathrm{si}$ can always be
represented as a pure state $\ket{\Psi}$ in some ``larger'' space
\cite{nielsen:book}. In our case, $\ket{\Psi}$ can be represented as a
sum of orthogonal contributions with amplitudes given by a set of
``partial waveshapes'' $\Psi_{\mathrm{si}}^{\{w\}}(t_s,t_i)$,
describing the signal-idler state in each of the configurations
$\{w\}$ of the rest of that larger space. Every of
$\Psi_{\mathrm{si}}^{\{w\}}$ evolves fully independently from the
others, and obey the same Schr\"odinger equation
\refeqs{eq:main-eqs-si}{eq:main-eqs-p}. Therefore, $\rho_\mathrm{si}$
will be processed correctly if all of $\Psi_{\mathrm{si}}^{\{w\}}$
satisfy the slowness condition \refeq{eq:slow}.

Furthermore, using the fourth-order coherence function
$\Gamma^{(2,2)}$ \cite{mandel:book}
\begin{gather} \nonumber
  \Gamma^{(2,2)}(\tau_s,\tau_i) =
\tr{\left(\rho_\mathrm{si}a_s^\dag(t_s-\tau_s)a_i^\dag(t_i-\tau_i)a_s(t_s)a_i(t_i)\right)},
\label{eq:gamma} \end{gather}
\refeq{eq:slow} can be reformulated in an equivalent form:
\begin{equation} \label{eq:slow-gamma} \Gamma^{(2,2)}(\tau_s,\tau_i)
  \approx \Gamma^{(2,2)}(0,0) \end{equation} for
$|\tau_s|\lesssim\sigma$, $|\tau_i|\lesssim\sigma$. 
\refeq{eq:slow-gamma} obviously can not be satisfied if
$\mathcal T^{(4)} \lesssim \sigma$, where $\mathcal T^{(4)}$ is the
coherence time defined by the width of
$\Gamma^{(2,2)}(\tau_s,\tau_i)$, so the opposite condition,
\begin{equation}
\mathcal T^{(4)} \gg \sigma, \label{eq:coh-time} 
\end{equation}
is therefore necessary for the validity of \refeq{eq:slow-gamma}. Except
some pathological waveshapes, $\Gamma^{(2,2)}$ decreases monotonously
near the origin $(\tau_s,\tau_i)=(0,0)$. Thus, in most cases,
$\mathcal T^{(4)} \gg \sigma$ delivers also the
sufficient condition and can be therefore considered as a suitable
criterion for the proper gate operation.

\textit{Discussion and conclusions.} We showed that CPC-based gates
can be "waveshape tolerant", that is, successfully process arbitrary
waveshapes keeping them intact, if the waveshapes vary slowly on the
timescale of
$\sigma$, the noninstantaneous response time of the nonlinearity. This
condition can be reformulated in the form of a constraint on the
fourth order coherence given by \refeq{eq:slow-gamma}, or, in the most
cases, as an even simpler condition \refeq{eq:coh-time} on the
corresponding coherence time, both applicable also to mixed states.
Importantly, since the pulses cross each other, also temporal offsets
are processed correctly if the available propagation length is large
enough (see Supplementary for more details on this).

The smallest $\sigma$ appear for off-resonant nonlinearities such as
in fused silica at optical frequencies, in which case it lies in
the attosecond range \cite{hofmann15,sommer16}. Our simulations indicate that
in this situation the gate operation is not impossible, if a
strong coherent pump in the TW/cm$^2$ range and pulse durations in the
femtosecond range are used, assuming that the waveguide dispersion is
managed properly and the frequency channels are well separated.

We remark that the slowness condition mentioned above is a very strong
property. Most waveshapes changing slowly on the scale of the working
wavelength $\lambda$ and thus on the scale of $\sigma$, since most
typically $\sigma\lesssim \lambda$ in optics. It is thus notoriously
difficult to produce experimentally a waveshape which breaks this
condition. In more complex logical circuits, limitation due to linear
optical elements become even more strict, since even ideal linear
optical elements are known to preserve only symmetric distributions
(like $\Psi_{\mathrm{si}}(t_s,t_i)=\Psi_{\mathrm{si}}(t_i,t_s)$)
\cite{uren03}.

Waveshape tolerantness relaxes significantly the requirements to the photon
sources needed for scalable computing. The photons do not need to be
indistinguishable, unentangled or even pure anymore. Moreover, as we
have seen, some entangled distributions promise an significant
advantage in the resulting gate size over the uncorrelated photons.

\begin{acknowledgments}
  I.B. and U.M. thank Deutsche Forschungsgemeinschaft (DFG, German
  Research Foundation), projects BA 4156/4-2 and MO 850-19/2 for
  support. I.B., A.D., M.K., and U.M. acknowledge support from
  Germany’s Excellence Strategy within the Cluster of Excellence EXC
  2122 PhoenixD (Project ID 390833453) and Germany’s Excellence
  Strategy EXC-2123 QuantumFrontiers (Project ID 390837967). M.K
  acknowledges support from the German ministry of education and
  research (PQuMAL project).
\end{acknowledgments}


\begin{thebibliography}{38}%
\makeatletter
\providecommand \@ifxundefined [1]{%
 \@ifx{#1\undefined}
}%
\providecommand \@ifnum [1]{%
 \ifnum #1\expandafter \@firstoftwo
 \else \expandafter \@secondoftwo
 \fi
}%
\providecommand \@ifx [1]{%
 \ifx #1\expandafter \@firstoftwo
 \else \expandafter \@secondoftwo
 \fi
}%
\providecommand \natexlab [1]{#1}%
\providecommand \enquote  [1]{``#1''}%
\providecommand \bibnamefont  [1]{#1}%
\providecommand \bibfnamefont [1]{#1}%
\providecommand \citenamefont [1]{#1}%
\providecommand \href@noop [0]{\@secondoftwo}%
\providecommand \href [0]{\begingroup \@sanitize@url \@href}%
\providecommand \@href[1]{\@@startlink{#1}\@@href}%
\providecommand \@@href[1]{\endgroup#1\@@endlink}%
\providecommand \@sanitize@url [0]{\catcode `\\12\catcode `\$12\catcode
  `\&12\catcode `\#12\catcode `\^12\catcode `\_12\catcode `\%12\relax}%
\providecommand \@@startlink[1]{}%
\providecommand \@@endlink[0]{}%
\providecommand \url  [0]{\begingroup\@sanitize@url \@url }%
\providecommand \@url [1]{\endgroup\@href {#1}{\urlprefix }}%
\providecommand \urlprefix  [0]{URL }%
\providecommand \Eprint [0]{\href }%
\providecommand \doibase [0]{https://doi.org/}%
\providecommand \selectlanguage [0]{\@gobble}%
\providecommand \bibinfo  [0]{\@secondoftwo}%
\providecommand \bibfield  [0]{\@secondoftwo}%
\providecommand \translation [1]{[#1]}%
\providecommand \BibitemOpen [0]{}%
\providecommand \bibitemStop [0]{}%
\providecommand \bibitemNoStop [0]{.\EOS\space}%
\providecommand \EOS [0]{\spacefactor3000\relax}%
\providecommand \BibitemShut  [1]{\csname bibitem#1\endcsname}%
\let\auto@bib@innerbib\@empty
\bibitem [{\citenamefont {Langford}\ \emph {et~al.}(2011)\citenamefont
  {Langford}, \citenamefont {Ramelow}, \citenamefont {Prevedel}, \citenamefont
  {Munro}, \citenamefont {Milburn},\ and\ \citenamefont
  {Zeilinger}}]{langford11}%
  \BibitemOpen
  \bibfield  {author} {\bibinfo {author} {\bibfnamefont {N.~K.}\ \bibnamefont
  {Langford}}, \bibinfo {author} {\bibfnamefont {S.}~\bibnamefont {Ramelow}},
  \bibinfo {author} {\bibfnamefont {R.}~\bibnamefont {Prevedel}}, \bibinfo
  {author} {\bibfnamefont {W.~J.}\ \bibnamefont {Munro}}, \bibinfo {author}
  {\bibfnamefont {G.~J.}\ \bibnamefont {Milburn}},\ and\ \bibinfo {author}
  {\bibfnamefont {A.}~\bibnamefont {Zeilinger}},\ }\bibfield  {title} {\bibinfo
  {title} {Efficient quantum computing using coherent photon conversion},\
  }\href {https://doi.org/10.1038/nature10463} {\bibfield  {journal} {\bibinfo
  {journal} {Nature}\ }\textbf {\bibinfo {volume} {478}},\ \bibinfo {pages}
  {360} (\bibinfo {year} {2011})}\BibitemShut {NoStop}%
\bibitem [{\citenamefont {Xia}\ \emph {et~al.}(2016)\citenamefont {Xia},
  \citenamefont {Johnsson}, \citenamefont {Knight}, \citenamefont {Twamley}
  \emph {et~al.}}]{xia16}%
  \BibitemOpen
  \bibfield  {author} {\bibinfo {author} {\bibfnamefont {K.}~\bibnamefont
  {Xia}}, \bibinfo {author} {\bibfnamefont {M.}~\bibnamefont {Johnsson}},
  \bibinfo {author} {\bibfnamefont {P.~L.}\ \bibnamefont {Knight}}, \bibinfo
  {author} {\bibfnamefont {J.}~\bibnamefont {Twamley}},\
  }\bibfield  {title} {\bibinfo {title} {Cavity-free scheme for nondestructive
  detection of a single optical photon},\ }\href@noop {} {\bibfield  {journal}
  {\bibinfo  {journal} {Phys. Rev. Lett.}\ }\textbf {\bibinfo {volume}
  {116}},\ \bibinfo {pages} {023601} (\bibinfo {year} {2016})}\BibitemShut
  {NoStop}%
\bibitem [{\citenamefont {Niu}\ \emph {et~al.}(2018)\citenamefont {Niu},
  \citenamefont {Chuang},\ and\ \citenamefont {Shapiro}}]{niu18}%
  \BibitemOpen
  \bibfield  {author} {\bibinfo {author} {\bibfnamefont {M.~Y.}\ \bibnamefont
  {Niu}}, \bibinfo {author} {\bibfnamefont {I.~L.}\ \bibnamefont {Chuang}},\
  and\ \bibinfo {author} {\bibfnamefont {J.~H.}\ \bibnamefont {Shapiro}},\
  }\bibfield  {title} {\bibinfo {title} {Qudit-basis universal quantum
  computation using $\chi^{(2)}$ interactions},\ }\href@noop {} {\bibfield
  {journal} {\bibinfo  {journal} {Phys. Rev. Lett.}\ }\textbf {\bibinfo
  {volume} {120}},\ \bibinfo {pages} {160502} (\bibinfo {year}
  {2018})}\BibitemShut {NoStop}%
\bibitem [{\citenamefont {Viswanathan}\ and\ \citenamefont
  {Gea-Banacloche}(2015)}]{viswanathan15}%
  \BibitemOpen
  \bibfield  {author} {\bibinfo {author} {\bibfnamefont {B.}~\bibnamefont
  {Viswanathan}}\ and\ \bibinfo {author} {\bibfnamefont {J.}~\bibnamefont
  {Gea-Banacloche}},\ }\bibfield  {title} {\bibinfo {title} {Multimode analysis
  of a conditional phase gate based on second-order nonlinearity},\ }\href@noop
  {} {\bibfield  {journal} {\bibinfo  {journal} {Phys. Rev. A}\ }\textbf
  {\bibinfo {volume} {92}},\ \bibinfo {pages} {042330} (\bibinfo {year}
  {2015})}\BibitemShut {NoStop}%
\bibitem [{\citenamefont {Viswanathan}\ and\ \citenamefont
  {Gea-Banacloche}(2018)}]{viswanathan18}%
  \BibitemOpen
  \bibfield  {author} {\bibinfo {author} {\bibfnamefont {B.}~\bibnamefont
  {Viswanathan}}\ and\ \bibinfo {author} {\bibfnamefont {J.}~\bibnamefont
  {Gea-Banacloche}},\ }\bibfield  {title} {\bibinfo {title} {Analytical results
  for a conditional phase shift between single-photon pulses in a nonlocal
  nonlinear medium},\ }\href@noop {} {\bibfield  {journal} {\bibinfo  {journal}
  {Phys. Rev. A}\ }\textbf {\bibinfo {volume} {97}},\ \bibinfo {pages}
  {032314} (\bibinfo {year} {2018})}\BibitemShut {NoStop}%
\bibitem [{\citenamefont {Kounalakis}\ \emph {et~al.}(2018)\citenamefont
  {Kounalakis}, \citenamefont {Dickel}, \citenamefont {Bruno}, \citenamefont
  {Langford},\ and\ \citenamefont
  {Steele}}]{kounalakis18-cross-kerr-superconduct}%
  \BibitemOpen
  \bibfield  {author} {\bibinfo {author} {\bibfnamefont {M.}~\bibnamefont
  {Kounalakis}}, \bibinfo {author} {\bibfnamefont {C.}~\bibnamefont {Dickel}},
  \bibinfo {author} {\bibfnamefont {A.}~\bibnamefont {Bruno}}, \bibinfo
  {author} {\bibfnamefont {N.~K.}\ \bibnamefont {Langford}},\ and\ \bibinfo
  {author} {\bibfnamefont {G.~A.}\ \bibnamefont {Steele}},\ }\bibfield  {title}
  {\bibinfo {title} {Tuneable hopping and nonlinear cross-kerr interactions in
  a high-coherence superconducting circuit},\ }\href@noop {} {\bibfield
  {journal} {\bibinfo  {journal} {npj Quantum Inf.}\ }\textbf {\bibinfo
  {volume} {4}},\ \bibinfo {pages} {1} (\bibinfo {year} {2018})}\BibitemShut
  {NoStop}%
\bibitem [{\citenamefont {Tiarks}\ \emph {et~al.}(2019)\citenamefont {Tiarks},
  \citenamefont {Schmidt-Eberle}, \citenamefont {Stolz}, \citenamefont
  {Rempe},\ and\ \citenamefont {D{\"u}rr}}]{tiarks19-rydberg}%
  \BibitemOpen
  \bibfield  {author} {\bibinfo {author} {\bibfnamefont {D.}~\bibnamefont
  {Tiarks}}, \bibinfo {author} {\bibfnamefont {S.}~\bibnamefont
  {Schmidt-Eberle}}, \bibinfo {author} {\bibfnamefont {T.}~\bibnamefont
  {Stolz}}, \bibinfo {author} {\bibfnamefont {G.}~\bibnamefont {Rempe}},\ and\
  \bibinfo {author} {\bibfnamefont {S.}~\bibnamefont {D{\"u}rr}},\ }\bibfield
  {title} {\bibinfo {title} {A photon--photon quantum gate based on rydberg
  interactions},\ }\href@noop {} {\bibfield  {journal} {\bibinfo  {journal}
  {Nature Physics}\ }\textbf {\bibinfo {volume} {15}},\ \bibinfo {pages} {124}
  (\bibinfo {year} {2019})}\BibitemShut {NoStop}%
\bibitem [{\citenamefont {Sagona-Stophel}\ \emph {et~al.}(2020)\citenamefont
  {Sagona-Stophel}, \citenamefont {Shahrokhshahi}, \citenamefont {Jordaan},
  \citenamefont {Namazi},\ and\ \citenamefont {Figueroa}}]{sagona-stophel20}%
  \BibitemOpen
  \bibfield  {author} {\bibinfo {author} {\bibfnamefont {S.}~\bibnamefont
  {Sagona-Stophel}}, \bibinfo {author} {\bibfnamefont {R.}~\bibnamefont
  {Shahrokhshahi}}, \bibinfo {author} {\bibfnamefont {B.}~\bibnamefont
  {Jordaan}}, \bibinfo {author} {\bibfnamefont {M.}~\bibnamefont {Namazi}},\
  and\ \bibinfo {author} {\bibfnamefont {E.}~\bibnamefont {Figueroa}},\
  }\bibfield  {title} {\bibinfo {title} {Conditional $\ensuremath{\pi}$-phase
  shift of single-photon-level pulses at room temperature},\ }\href
  {https://doi.org/10.1103/PhysRevLett.125.243601} {\bibfield  {journal}
  {\bibinfo  {journal} {Phys. Rev. Lett.}\ }\textbf {\bibinfo {volume} {125}},\
  \bibinfo {pages} {243601} (\bibinfo {year} {2020})}\BibitemShut {NoStop}%
\bibitem [{\citenamefont {Heuck}\ \emph {et~al.}(2020)\citenamefont {Heuck},
  \citenamefont {Jacobs},\ and\ \citenamefont {Englund}}]{heuck20}%
  \BibitemOpen
  \bibfield  {author} {\bibinfo {author} {\bibfnamefont {M.}~\bibnamefont
  {Heuck}}, \bibinfo {author} {\bibfnamefont {K.}~\bibnamefont {Jacobs}},\ and\
  \bibinfo {author} {\bibfnamefont {D.~R.}\ \bibnamefont {Englund}},\
  }\bibfield  {title} {\bibinfo {title} {Controlled-phase gate using
  dynamically coupled cavities and optical nonlinearities},\ }\href@noop {}
  {\bibfield  {journal} {\bibinfo  {journal} {Phys. Rev. Lett.}\
  }\textbf {\bibinfo {volume} {124}},\ \bibinfo {pages} {160501} (\bibinfo
  {year} {2020})}\BibitemShut {NoStop}%
\bibitem [{\citenamefont {Dot}\ \emph {et~al.}(2014)\citenamefont {Dot},
  \citenamefont {Meyer-Scott}, \citenamefont {Ahmad}, \citenamefont
  {Rochette},\ and\ \citenamefont {Jennewein}}]{dot14}%
  \BibitemOpen
  \bibfield  {author} {\bibinfo {author} {\bibfnamefont {Audrey}\ \bibnamefont
  {Dot}}, \bibinfo {author} {\bibfnamefont {Evan}\ \bibnamefont {Meyer-Scott}},
  \bibinfo {author} {\bibfnamefont {Raja}\ \bibnamefont {Ahmad}}, \bibinfo
  {author} {\bibfnamefont {Martin}\ \bibnamefont {Rochette}}, \ and\ \bibinfo
  {author} {\bibfnamefont {Thomas}\ \bibnamefont {Jennewein}},\ }\bibfield
  {title} {\enquote {\bibinfo {title} {Converting one photon into two via
  four-wave mixing in optical fibers},}\ }\href {\doibase
  10.1103/PhysRevA.90.043808} {\bibfield  {journal} {\bibinfo  {journal} {Phys.
  Rev. A}\ }\textbf {\bibinfo {volume} {90}},\ \bibinfo {pages} {043808}
  (\bibinfo {year} {2014})}\BibitemShut {NoStop}%
\bibitem [{\citenamefont {Meyer-Scott}\ \emph {et~al.}(2015)\citenamefont
  {Meyer-Scott}, \citenamefont {Dot}, \citenamefont {Ahmad}, \citenamefont
  {Li}, \citenamefont {Rochette},\ and\ \citenamefont
  {Jennewein}}]{meyer-scott15}%
  \BibitemOpen
  \bibfield  {author} {\bibinfo {author} {\bibfnamefont {Evan}\ \bibnamefont
  {Meyer-Scott}}, \bibinfo {author} {\bibfnamefont {Audrey}\ \bibnamefont
  {Dot}}, \bibinfo {author} {\bibfnamefont {Raja}\ \bibnamefont {Ahmad}},
  \bibinfo {author} {\bibfnamefont {Lizhu}\ \bibnamefont {Li}}, \bibinfo
  {author} {\bibfnamefont {Martin}\ \bibnamefont {Rochette}}, \ and\ \bibinfo
  {author} {\bibfnamefont {Thomas}\ \bibnamefont {Jennewein}},\ }\bibfield
  {title} {\enquote {\bibinfo {title} {Power-efficient production of photon
  pairs in a tapered chalcogenide microwire},}\ }\href {\doibase
  10.1063/1.4913743} {\bibfield  {journal} {\bibinfo  {journal} {Applied
  Physics Letters}\ }\textbf {\bibinfo {volume} {106}},\ \bibinfo {pages}
  {081111} (\bibinfo {year} {2015})}\BibitemShut {NoStop}%
\bibitem [{\citenamefont {Solntsev}\ \emph {et~al.}(2021)\citenamefont
  {Solntsev}, \citenamefont {Batalov}, \citenamefont {Langford},\ and\
  \citenamefont {Sukhorukov}}]{solntsev21}%
  \BibitemOpen
  \bibfield  {author} {\bibinfo {author} {\bibfnamefont {Alexander~S.}\
  \bibnamefont {Solntsev}}, \bibinfo {author} {\bibfnamefont {Sergey~V.}\
  \bibnamefont {Batalov}}, \bibinfo {author} {\bibfnamefont {Nathan~K.}\
  \bibnamefont {Langford}}, \ and\ \bibinfo {author} {\bibfnamefont
  {Andrey~A.}\ \bibnamefont {Sukhorukov}},\ }\href@noop {} {\enquote {\bibinfo
  {title} {Complete conversion between one and two photons in nonlinear
  waveguides with tailored dispersion},}\ } (\bibinfo {year} {2021}),\ \Eprint
  {http://arxiv.org/abs/2110.03110} {arXiv:2110.03110 [physics.optics]}
  \BibitemShut {NoStop}%
\bibitem [{\citenamefont {Knill}\ \emph {et~al.}(2001)\citenamefont {Knill},
  \citenamefont {Laflamme},\ and\ \citenamefont {Milburn}}]{knill01}%
  \BibitemOpen
  \bibfield  {author} {\bibinfo {author} {\bibfnamefont {E.}~\bibnamefont
  {Knill}}, \bibinfo {author} {\bibfnamefont {R.}~\bibnamefont {Laflamme}},\
  and\ \bibinfo {author} {\bibfnamefont {G.~J.}\ \bibnamefont {Milburn}},\
  }\bibfield  {title} {\bibinfo {title} {A scheme for efficient quantum
  computation with linear optics},\ }\href {https://doi.org/10.1038/35051009}
  {\bibfield  {journal} {\bibinfo  {journal} {Nature}\ }\textbf {\bibinfo
  {volume} {409}},\ \bibinfo {pages} {46} (\bibinfo {year} {2001})}\BibitemShut
  {NoStop}%
\bibitem [{\citenamefont {O'Brien}\ \emph {et~al.}(2003)\citenamefont
  {O'Brien}, \citenamefont {Pryde}, \citenamefont {White}, \citenamefont
  {Ralph},\ and\ \citenamefont {Branning}}]{obrien03}%
  \BibitemOpen
  \bibfield  {author} {\bibinfo {author} {\bibfnamefont {J.~L.}\ \bibnamefont
  {O'Brien}}, \bibinfo {author} {\bibfnamefont {G.~J.}\ \bibnamefont {Pryde}},
  \bibinfo {author} {\bibfnamefont {A.~G.}\ \bibnamefont {White}}, \bibinfo
  {author} {\bibfnamefont {T.~C.}\ \bibnamefont {Ralph}},\ and\ \bibinfo
  {author} {\bibfnamefont {D.}~\bibnamefont {Branning}},\ }\bibfield  {title}
  {\bibinfo {title} {Demonstration of an all-optical quantum controlled-not
  gate},\ }\href {https://doi.org/10.1038/nature02054} {\bibfield  {journal}
  {\bibinfo  {journal} {Nature}\ }\textbf {\bibinfo {volume} {426}},\ \bibinfo
  {pages} {264} (\bibinfo {year} {2003})}\BibitemShut {NoStop}%
\bibitem [{\citenamefont {Kok}\ \emph {et~al.}(2007)\citenamefont {Kok},
  \citenamefont {Munro}, \citenamefont {Nemoto}, \citenamefont {Ralph},
  \citenamefont {Dowling},\ and\ \citenamefont {Milburn}}]{kok07-rev}%
  \BibitemOpen
  \bibfield  {author} {\bibinfo {author} {\bibfnamefont {P.}~\bibnamefont
  {Kok}}, \bibinfo {author} {\bibfnamefont {W.~J.}\ \bibnamefont {Munro}},
  \bibinfo {author} {\bibfnamefont {K.}~\bibnamefont {Nemoto}}, \bibinfo
  {author} {\bibfnamefont {T.~C.}\ \bibnamefont {Ralph}}, \bibinfo {author}
  {\bibfnamefont {J.~P.}\ \bibnamefont {Dowling}},\ and\ \bibinfo {author}
  {\bibfnamefont {G.~J.}\ \bibnamefont {Milburn}},\ }\bibfield  {title}
  {\bibinfo {title} {Linear optical quantum computing with photonic qubits},\
  }\href {http://link.aps.org/abstract/RMP/v79/p135} {\bibfield  {journal}
  {\bibinfo  {journal} {Rev. Mod. Phys.}\ }\textbf {\bibinfo {volume} {79}},\
  \bibinfo {pages} {135} (\bibinfo {year} {2007})}\BibitemShut {NoStop}%
\bibitem [{\citenamefont {Carolan}\ \emph {et~al.}(2015)\citenamefont
  {Carolan}, \citenamefont {Harrold}, \citenamefont {Sparrow}, \citenamefont
  {Mart{\'\i}n-L{\'o}pez}, \citenamefont {Russell}, \citenamefont
  {Silverstone}, \citenamefont {Shadbolt}, \citenamefont {Matsuda},
  \citenamefont {Oguma}, \citenamefont {Itoh}, \citenamefont {Marshall},
  \citenamefont {Thompson}, \citenamefont {Matthews}, \citenamefont
  {Hashimoto}, \citenamefont {O{\textquoteright}Brien},\ and\ \citenamefont
  {Laing}}]{carolan15}%
  \BibitemOpen
  \bibfield  {author} {\bibinfo {author} {\bibfnamefont {J.}~\bibnamefont
  {Carolan}}, \bibinfo {author} {\bibfnamefont {C.}~\bibnamefont {Harrold}},
  \bibinfo {author} {\bibfnamefont {C.}~\bibnamefont {Sparrow}}, \bibinfo
  {author} {\bibfnamefont {E.}~\bibnamefont {Mart{\'\i}n-L{\'o}pez}}, \bibinfo
  {author} {\bibfnamefont {N.~J.}\ \bibnamefont {Russell}}, \bibinfo {author}
  {\bibfnamefont {J.~W.}\ \bibnamefont {Silverstone}}, \bibinfo {author}
  {\bibfnamefont {P.~J.}\ \bibnamefont {Shadbolt}}, \bibinfo {author}
  {\bibfnamefont {N.}~\bibnamefont {Matsuda}}, \bibinfo {author} {\bibfnamefont
  {M.}~\bibnamefont {Oguma}}, \bibinfo {author} {\bibfnamefont
  {M.}~\bibnamefont {Itoh}}, \bibinfo {author} {\bibfnamefont {G.~D.}\
  \bibnamefont {Marshall}}, \bibinfo {author} {\bibfnamefont {M.~G.}\
  \bibnamefont {Thompson}}, \bibinfo {author} {\bibfnamefont {J.~C.~F.}\
  \bibnamefont {Matthews}}, \bibinfo {author} {\bibfnamefont {T.}~\bibnamefont
  {Hashimoto}}, \bibinfo {author} {\bibfnamefont {J.~L.}\ \bibnamefont
  {O{\textquoteright}Brien}},\ and\ \bibinfo {author} {\bibfnamefont
  {A.}~\bibnamefont {Laing}},\ }\bibfield  {title} {\bibinfo {title} {Universal
  linear optics},\ }\href {https://doi.org/10.1126/science.aab3642} {\bibfield
  {journal} {\bibinfo  {journal} {Science}\ }\textbf {\bibinfo {volume}
  {349}},\ \bibinfo {pages} {711} (\bibinfo {year} {2015})}\BibitemShut
  {NoStop}%
\bibitem [{\citenamefont {Marcikic}\ \emph {et~al.}(2002)\citenamefont
  {Marcikic}, \citenamefont {de~Riedmatten}, \citenamefont {Tittel},
  \citenamefont {Scarani}, \citenamefont {Zbinden},\ and\ \citenamefont
  {Gisin}}]{marcikic02-time-bins}%
  \BibitemOpen
  \bibfield  {author} {\bibinfo {author} {\bibfnamefont {I.}~\bibnamefont
  {Marcikic}}, \bibinfo {author} {\bibfnamefont {H.}~\bibnamefont
  {de~Riedmatten}}, \bibinfo {author} {\bibfnamefont {W.}~\bibnamefont
  {Tittel}}, \bibinfo {author} {\bibfnamefont {V.}~\bibnamefont {Scarani}},
  \bibinfo {author} {\bibfnamefont {H.}~\bibnamefont {Zbinden}},\ and\ \bibinfo
  {author} {\bibfnamefont {N.}~\bibnamefont {Gisin}},\ }\bibfield  {title}
  {\bibinfo {title} {Time-bin entangled qubits for quantum communication
  created by femtosecond pulses},\ }\href
  {https://doi.org/10.1103/PhysRevA.66.062308} {\bibfield  {journal} {\bibinfo
  {journal} {Phys. Rev. A}\ }\textbf {\bibinfo {volume} {66}},\ \bibinfo
  {pages} {062308} (\bibinfo {year} {2002})}\BibitemShut {NoStop}%
\bibitem [{\citenamefont {Xiong}\ \emph {et~al.}(2015)\citenamefont {Xiong},
  \citenamefont {Zhang}, \citenamefont {Mahendra}, \citenamefont {He},
  \citenamefont {Choi}, \citenamefont {Chae}, \citenamefont {Marpaung},
  \citenamefont {Leinse}, \citenamefont {Heideman}, \citenamefont {Hoekman},
  \citenamefont {Roeloffzen}, \citenamefont {Oldenbeuving}, \citenamefont {van
  Dijk}, \citenamefont {Taddei}, \citenamefont {Leong},\ and\ \citenamefont
  {Eggleton}}]{xiong15-time-bin-exp}%
  \BibitemOpen
  \bibfield  {author} {\bibinfo {author} {\bibfnamefont {C.}~\bibnamefont
  {Xiong}}, \bibinfo {author} {\bibfnamefont {X.}~\bibnamefont {Zhang}},
  \bibinfo {author} {\bibfnamefont {A.}~\bibnamefont {Mahendra}}, \bibinfo
  {author} {\bibfnamefont {J.}~\bibnamefont {He}}, \bibinfo {author}
  {\bibfnamefont {D.-Y.}\ \bibnamefont {Choi}}, \bibinfo {author}
  {\bibfnamefont {C.~J.}\ \bibnamefont {Chae}}, \bibinfo {author}
  {\bibfnamefont {D.}~\bibnamefont {Marpaung}}, \bibinfo {author}
  {\bibfnamefont {A.}~\bibnamefont {Leinse}}, \bibinfo {author} {\bibfnamefont
  {R.~G.}\ \bibnamefont {Heideman}}, \bibinfo {author} {\bibfnamefont
  {M.}~\bibnamefont {Hoekman}}, \bibinfo {author} {\bibfnamefont {C.~G.~H.}\
  \bibnamefont {Roeloffzen}}, \bibinfo {author} {\bibfnamefont {R.~M.}\
  \bibnamefont {Oldenbeuving}}, \bibinfo {author} {\bibfnamefont {P.~W.~L.}\
  \bibnamefont {van Dijk}}, \bibinfo {author} {\bibfnamefont {C.}~\bibnamefont
  {Taddei}}, \bibinfo {author} {\bibfnamefont {P.~H.~W.}\ \bibnamefont
  {Leong}},\ and\ \bibinfo {author} {\bibfnamefont {B.~J.}\ \bibnamefont
  {Eggleton}},\ }\bibfield  {title} {\bibinfo {title} {Compact and
  reconfigurable silicon nitride time-bin entanglement circuit},\ }\href
  {https://doi.org/10.1364/OPTICA.2.000724} {\bibfield  {journal} {\bibinfo
  {journal} {Optica}\ }\textbf {\bibinfo {volume} {2}},\ \bibinfo {pages} {724}
  (\bibinfo {year} {2015})}\BibitemShut {NoStop}%
\bibitem [{\citenamefont {Brecht}\ \emph {et~al.}(2015)\citenamefont {Brecht},
  \citenamefont {Reddy}, \citenamefont {Silberhorn},\ and\ \citenamefont
  {Raymer}}]{brecht15}%
  \BibitemOpen
  \bibfield  {author} {\bibinfo {author} {\bibfnamefont {B.}~\bibnamefont
  {Brecht}}, \bibinfo {author} {\bibfnamefont {D.~V.}\ \bibnamefont {Reddy}},
  \bibinfo {author} {\bibfnamefont {C.}~\bibnamefont {Silberhorn}},\ and\
  \bibinfo {author} {\bibfnamefont {M.~G.}~\bibnamefont {Raymer}},\ }\bibfield
  {title} {\bibinfo {title} {Photon temporal modes: a complete framework for
  quantum information science},\ }\href@noop {} {\bibfield  {journal} {\bibinfo
   {journal} {Phys. Rev. X}\ }\textbf {\bibinfo {volume} {5}},\ \bibinfo
  {pages} {041017} (\bibinfo {year} {2015})}\BibitemShut {NoStop}%
\bibitem [{\citenamefont {Ansari}\ \emph {et~al.}(2018)\citenamefont {Ansari},
  \citenamefont {Donohue}, \citenamefont {Brecht},\ and\ \citenamefont
  {Silberhorn}}]{ansari18-temporal-modes}%
  \BibitemOpen
  \bibfield  {author} {\bibinfo {author} {\bibfnamefont {V.}~\bibnamefont
  {Ansari}}, \bibinfo {author} {\bibfnamefont {J.~M.}\ \bibnamefont {Donohue}},
  \bibinfo {author} {\bibfnamefont {B.}~\bibnamefont {Brecht}},\ and\ \bibinfo
  {author} {\bibfnamefont {C.}~\bibnamefont {Silberhorn}},\ }\bibfield  {title}
  {\bibinfo {title} {Tailoring nonlinear processes for quantum optics with
  pulsed temporal-mode encodings},\ }\href@noop {} {\bibfield  {journal}
  {\bibinfo  {journal} {Optica}\ }\textbf {\bibinfo {volume} {5}},\ \bibinfo
  {pages} {534} (\bibinfo {year} {2018})}\BibitemShut {NoStop}%
\bibitem [{\citenamefont {Luo}\ \emph {et~al.}(2019)\citenamefont {Luo},
  \citenamefont {Brauner}, \citenamefont {Eigner}, \citenamefont {Sharapova},
  \citenamefont {Ricken}, \citenamefont {Meier}, \citenamefont {Herrmann},\
  and\ \citenamefont {Silberhorn}}]{luo19-electroopt-circuits}%
  \BibitemOpen
  \bibfield  {author} {\bibinfo {author} {\bibfnamefont {K.-H.}\ \bibnamefont
  {Luo}}, \bibinfo {author} {\bibfnamefont {S.}~\bibnamefont {Brauner}},
  \bibinfo {author} {\bibfnamefont {C.}~\bibnamefont {Eigner}}, \bibinfo
  {author} {\bibfnamefont {P.~R.}\ \bibnamefont {Sharapova}}, \bibinfo {author}
  {\bibfnamefont {R.}~\bibnamefont {Ricken}}, \bibinfo {author} {\bibfnamefont
  {T.}~\bibnamefont {Meier}}, \bibinfo {author} {\bibfnamefont
  {H.}~\bibnamefont {Herrmann}},\ and\ \bibinfo {author} {\bibfnamefont
  {C.}~\bibnamefont {Silberhorn}},\ }\bibfield  {title} {\bibinfo {title}
  {Nonlinear integrated quantum electro-optic circuits},\ }\bibfield  {journal}
  {\bibinfo  {journal} {Science Advances}\ }\textbf {\bibinfo {volume} {5}},\
  \href {https://doi.org/10.1126/sciadv.aat1451} {10.1126/sciadv.aat1451}
  (\bibinfo {year} {2019})\BibitemShut {NoStop}%
\bibitem [{\citenamefont {Ramelow}\ \emph {et~al.}(2009)\citenamefont
  {Ramelow}, \citenamefont {Ratschbacher}, \citenamefont {Fedrizzi},
  \citenamefont {Langford},\ and\ \citenamefont
  {Zeilinger}}]{ramelow09-freq-bins}%
  \BibitemOpen
  \bibfield  {author} {\bibinfo {author} {\bibfnamefont {S.}~\bibnamefont
  {Ramelow}}, \bibinfo {author} {\bibfnamefont {L.}~\bibnamefont
  {Ratschbacher}}, \bibinfo {author} {\bibfnamefont {A.}~\bibnamefont
  {Fedrizzi}}, \bibinfo {author} {\bibfnamefont {N.~K.}~\bibnamefont {Langford}},\
  and\ \bibinfo {author} {\bibfnamefont {A.}~\bibnamefont {Zeilinger}},\
  }\bibfield  {title} {\bibinfo {title} {Discrete tunable color entanglement},\
  }\href@noop {} {\bibfield  {journal} {\bibinfo  {journal} {Phys. Rev.
  Lett.}\ }\textbf {\bibinfo {volume} {103}},\ \bibinfo {pages} {253601}
  (\bibinfo {year} {2009})}\BibitemShut {NoStop}%
\bibitem [{\citenamefont {Olislager}\ \emph {et~al.}(2010)\citenamefont
  {Olislager}, \citenamefont {Cussey}, \citenamefont {Nguyen}, \citenamefont
  {Emplit}, \citenamefont {Massar}, \citenamefont {Merolla},\ and\
  \citenamefont {Huy}}]{olislager10-frequency-bins}%
  \BibitemOpen
  \bibfield  {author} {\bibinfo {author} {\bibfnamefont {L.}~\bibnamefont
  {Olislager}}, \bibinfo {author} {\bibfnamefont {J.}~\bibnamefont {Cussey}},
  \bibinfo {author} {\bibfnamefont {A.~T.}\ \bibnamefont {Nguyen}}, \bibinfo
  {author} {\bibfnamefont {P.}~\bibnamefont {Emplit}}, \bibinfo {author}
  {\bibfnamefont {S.}~\bibnamefont {Massar}}, \bibinfo {author} {\bibfnamefont
  {J.-M.}\ \bibnamefont {Merolla}},\ and\ \bibinfo {author} {\bibfnamefont
  {K.~P.}\ \bibnamefont {Huy}},\ }\bibfield  {title} {\bibinfo {title}
  {Frequency-bin entangled photons},\ }\href@noop {} {\bibfield  {journal}
  {\bibinfo  {journal} {Phys. Rev. A}\ }\textbf {\bibinfo {volume} {82}},\
  \bibinfo {pages} {013804} (\bibinfo {year} {2010})}\BibitemShut {NoStop}%
\bibitem [{\citenamefont {Reimer}\ \emph {et~al.}(2016)\citenamefont {Reimer},
  \citenamefont {Kues}, \citenamefont {Roztocki}, \citenamefont {Wetzel},
  \citenamefont {Grazioso}, \citenamefont {Little}, \citenamefont {Chu},
  \citenamefont {Johnston}, \citenamefont {Bromberg}, \citenamefont {Caspani},
  \citenamefont {Moss},\ and\ \citenamefont {Morandotti}}]{reimer16}%
  \BibitemOpen
  \bibfield  {author} {\bibinfo {author} {\bibfnamefont {C.}~\bibnamefont
  {Reimer}}, \bibinfo {author} {\bibfnamefont {M.}~\bibnamefont {Kues}},
  \bibinfo {author} {\bibfnamefont {P.}~\bibnamefont {Roztocki}}, \bibinfo
  {author} {\bibfnamefont {B.}~\bibnamefont {Wetzel}}, \bibinfo {author}
  {\bibfnamefont {F.}~\bibnamefont {Grazioso}}, \bibinfo {author}
  {\bibfnamefont {B.~E.}\ \bibnamefont {Little}}, \bibinfo {author}
  {\bibfnamefont {S.~T.}\ \bibnamefont {Chu}}, \bibinfo {author} {\bibfnamefont
  {T.}~\bibnamefont {Johnston}}, \bibinfo {author} {\bibfnamefont
  {Y.}~\bibnamefont {Bromberg}}, \bibinfo {author} {\bibfnamefont
  {L.}~\bibnamefont {Caspani}}, \bibinfo {author} {\bibfnamefont {D.~J.}\
  \bibnamefont {Moss}},\ and\ \bibinfo {author} {\bibfnamefont
  {R.}~\bibnamefont {Morandotti}},\ }\bibfield  {title} {\bibinfo {title}
  {Generation of multiphoton entangled quantum states by means of integrated
  frequency combs},\ }\href {https://doi.org/10.1126/science.aad8532}
  {\bibfield  {journal} {\bibinfo  {journal} {Science}\ }\textbf {\bibinfo
  {volume} {351}},\ \bibinfo {pages} {1176} (\bibinfo {year}
  {2016})}\BibitemShut {NoStop}%
\bibitem [{\citenamefont {Kues}\ \emph {et~al.}(2017)\citenamefont {Kues},
  \citenamefont {Reimer}, \citenamefont {Roztocki}, \citenamefont {Cort{\'e}s},
  \citenamefont {Sciara}, \citenamefont {Wetzel}, \citenamefont {Zhang},
  \citenamefont {Cino}, \citenamefont {Chu}, \citenamefont {Little} \emph
  {et~al.}}]{kues17}%
  \BibitemOpen
  \bibfield  {author} {\bibinfo {author} {\bibfnamefont {M.}~\bibnamefont
  {Kues}}, \bibinfo {author} {\bibfnamefont {C.}~\bibnamefont {Reimer}},
  \bibinfo {author} {\bibfnamefont {P.}~\bibnamefont {Roztocki}}, \bibinfo
  {author} {\bibfnamefont {L.~R.}\ \bibnamefont {Cort{\'e}s}}, \bibinfo
  {author} {\bibfnamefont {S.}~\bibnamefont {Sciara}}, \bibinfo {author}
  {\bibfnamefont {B.}~\bibnamefont {Wetzel}}, \bibinfo {author} {\bibfnamefont
  {Y.}~\bibnamefont {Zhang}}, \bibinfo {author} {\bibfnamefont
  {A.}~\bibnamefont {Cino}}, \bibinfo {author} {\bibfnamefont {S.~T.}\
  \bibnamefont {Chu}}, \bibinfo {author} {\bibfnamefont {B.~E.}\ \bibnamefont
  {Little}}, \emph {et~al.},\ }\bibfield  {title} {\bibinfo {title} {On-chip
  generation of high-dimensional entangled quantum states and their coherent
  control},\ }\href@noop {} {\bibfield  {journal} {\bibinfo  {journal}
  {Nature}\ }\textbf {\bibinfo {volume} {546}},\ \bibinfo {pages} {622}
  (\bibinfo {year} {2017})}\BibitemShut {NoStop}%
\bibitem [{\citenamefont {{Braunstein}}\ and\ \citenamefont {{van
  Loock}}(2005)}]{braunstein05}%
  \BibitemOpen
  \bibfield  {author} {\bibinfo {author} {\bibfnamefont {S.~L.}\ \bibnamefont
  {{Braunstein}}}\ and\ \bibinfo {author} {\bibfnamefont {P.}~\bibnamefont
  {{van Loock}}},\ }\bibfield  {title} {\bibinfo {title} {{Quantum information
  with continuous variables}},\ }\href
  {https://doi.org/10.1103/RevModPhys.77.513} {\bibfield  {journal} {\bibinfo
  {journal} {Rev. Mod. Phys}\ }\textbf {\bibinfo {volume} {77}},\
  \bibinfo {pages} {513} (\bibinfo {year} {2005})},\ \Eprint
  {https://arxiv.org/abs/quant-ph/0410100} {quant-ph/0410100} \BibitemShut
  {NoStop}%
\bibitem [{\citenamefont {Andersen}\ \emph {et~al.}(2010)\citenamefont
  {Andersen}, \citenamefont {Leuchs},\ and\ \citenamefont
  {Silberhorn}}]{andersen10-rev-cont-vars}%
  \BibitemOpen
  \bibfield  {author} {\bibinfo {author} {\bibfnamefont {U.}~\bibnamefont
  {Andersen}}, \bibinfo {author} {\bibfnamefont {G.}~\bibnamefont {Leuchs}},\
  and\ \bibinfo {author} {\bibfnamefont {C.}~\bibnamefont {Silberhorn}},\
  }\bibfield  {title} {\bibinfo {title} {Continuous-variable quantum
  information processing},\ }\href
  {https://doi.org/https://doi.org/10.1002/lpor.200910010} {\bibfield
  {journal} {\bibinfo  {journal} {Laser \& Photon. Rev.}\ }\textbf
  {\bibinfo {volume} {4}},\ \bibinfo {pages} {337} (\bibinfo {year}
  {2010})}\BibitemShut {NoStop}%
\bibitem [{\citenamefont {Mair}\ \emph {et~al.}(2001)\citenamefont {Mair},
  \citenamefont {Vaziri}, \citenamefont {Weihs},\ and\ \citenamefont
  {Zeilinger}}]{mair01-entanglement-oam}%
  \BibitemOpen
  \bibfield  {author} {\bibinfo {author} {\bibfnamefont {A.}~\bibnamefont
  {Mair}}, \bibinfo {author} {\bibfnamefont {A.}~\bibnamefont {Vaziri}},
  \bibinfo {author} {\bibfnamefont {G.}~\bibnamefont {Weihs}},\ and\ \bibinfo
  {author} {\bibfnamefont {A.}~\bibnamefont {Zeilinger}},\ }\bibfield  {title}
  {\bibinfo {title} {Entanglement of the orbital angular momentum states of
  photons},\ }\href@noop {} {\bibfield  {journal} {\bibinfo  {journal}
  {Nature}\ }\textbf {\bibinfo {volume} {412}},\ \bibinfo {pages} {313}
  (\bibinfo {year} {2001})}\BibitemShut {NoStop}%
\bibitem [{\citenamefont {U'ren}\ \emph {et~al.}(2003)\citenamefont {U'ren},
  \citenamefont {Banaszek},\ and\ \citenamefont {Walmsley}}]{uren03}%
  \BibitemOpen
  \bibfield  {author} {\bibinfo {author} {\bibfnamefont {A.~B.}\ \bibnamefont
  {U'ren}}, \bibinfo {author} {\bibfnamefont {K.}~\bibnamefont {Banaszek}},\
  and\ \bibinfo {author} {\bibfnamefont {I.~A.}\ \bibnamefont {Walmsley}},\
  }\bibfield  {title} {\bibinfo {title} {{Photon engineering for quantum
  information processing}},\ }\href {http://cds.cern.ch/record/618954}
  {\bibfield  {journal} {\bibinfo  {journal} {Quantum Inf. Comput.}\ }\textbf
  {\bibinfo {volume} {3}},\ \bibinfo {pages} {480. 23 p} (\bibinfo {year}
  {2003})}\BibitemShut {NoStop}%
\bibitem [{\citenamefont {U'Ren}\ \emph {et~al.}(2005)\citenamefont {U'Ren},
  \citenamefont {Silberhorn}, \citenamefont {Banaszek}, \citenamefont
  {Walmsley}, \citenamefont {Erdmann}, \citenamefont {Grice},\ and\
  \citenamefont {Raymer}}]{uren05}%
  \BibitemOpen
  \bibfield  {author} {\bibinfo {author} {\bibfnamefont {A.}~\bibnamefont
  {U'Ren}}, \bibinfo {author} {\bibfnamefont {C.}~\bibnamefont {Silberhorn}},
  \bibinfo {author} {\bibfnamefont {K.}~\bibnamefont {Banaszek}}, \bibinfo
  {author} {\bibfnamefont {I.}~\bibnamefont {Walmsley}}, \bibinfo {author}
  {\bibfnamefont {R.}~\bibnamefont {Erdmann}}, \bibinfo {author} {\bibfnamefont
  {W.}~\bibnamefont {Grice}},\ and\ \bibinfo {author} {\bibfnamefont
  {M.}~\bibnamefont {Raymer}},\ }\bibfield  {title} {\bibinfo {title}
  {Generation of pure-state single-photon wavepackets by conditional
  preparation based on spontaneous parametiric downconversion},\ }\href@noop {}
  {\bibfield  {journal} {\bibinfo  {journal} {Laser Phys.}\ }\textbf
  {\bibinfo {volume} {15}} (\bibinfo {year} {2005})}\BibitemShut {NoStop}%
\bibitem [{\citenamefont {Babushkin}\ \emph {et~al.}(2020)\citenamefont
  {Babushkin}, \citenamefont {Morgner},\ and\ \citenamefont
  {Demircan}}]{babushkin20}%
  \BibitemOpen
  \bibfield  {author} {\bibinfo {author} {\bibfnamefont {I.}~\bibnamefont
  {Babushkin}}, \bibinfo {author} {\bibfnamefont {U.}~\bibnamefont {Morgner}},\
  and\ \bibinfo {author} {\bibfnamefont {A.}~\bibnamefont {Demircan}},\
  }\bibfield  {title} {\bibinfo {title} {Stability of quantum linear logic
  circuits against perturbations},\ }\href@noop {} {\bibfield  {journal}
  {\bibinfo  {journal} {J. Phys. A: Math. Theor.}\
  }\textbf {\bibinfo {volume} {53}},\ \bibinfo {pages} {445307} (\bibinfo
  {year} {2020})}\BibitemShut {NoStop}%
\bibitem [{\citenamefont {Liang}\ \emph {et~al.}(2018)\citenamefont {Liang},
  \citenamefont {Venkatramani}, \citenamefont {Cantu}, \citenamefont
  {Nicholson}, \citenamefont {Gullans}, \citenamefont {Gorshkov}, \citenamefont
  {Thompson}, \citenamefont {Chin}, \citenamefont {Lukin},\ and\ \citenamefont
  {Vuleti{\'c}}}]{liang18-3photon}%
  \BibitemOpen
  \bibfield  {author} {\bibinfo {author} {\bibfnamefont {Q.-Y.}\ \bibnamefont
  {Liang}}, \bibinfo {author} {\bibfnamefont {A.~V.}\ \bibnamefont
  {Venkatramani}}, \bibinfo {author} {\bibfnamefont {S.~H.}\ \bibnamefont
  {Cantu}}, \bibinfo {author} {\bibfnamefont {T.~L.}\ \bibnamefont
  {Nicholson}}, \bibinfo {author} {\bibfnamefont {M.~J.}\ \bibnamefont
  {Gullans}}, \bibinfo {author} {\bibfnamefont {A.~V.}\ \bibnamefont
  {Gorshkov}}, \bibinfo {author} {\bibfnamefont {J.~D.}\ \bibnamefont
  {Thompson}}, \bibinfo {author} {\bibfnamefont {C.}~\bibnamefont {Chin}},
  \bibinfo {author} {\bibfnamefont {M.~D.}\ \bibnamefont {Lukin}},\ and\
  \bibinfo {author} {\bibfnamefont {V.}~\bibnamefont {Vuleti{\'c}}},\
  }\bibfield  {title} {\bibinfo {title} {Observation of three-photon bound
  states in a quantum nonlinear medium},\ }\href@noop {} {\bibfield  {journal}
  {\bibinfo  {journal} {Science}\ }\textbf {\bibinfo {volume} {359}},\ \bibinfo
  {pages} {783} (\bibinfo {year} {2018})}\BibitemShut {NoStop}%
\bibitem [{\citenamefont {Boyd}\ and\ \citenamefont
  {Boyd}(1992)}]{boyd92:book}%
  \BibitemOpen
  \bibfield  {author} {\bibinfo {author} {\bibfnamefont {R.}~\bibnamefont
  {Boyd}}\ and\ \bibinfo {author} {\bibfnamefont {M.}~\bibnamefont {Boyd}},\
  }\href {https://books.google.de/books?id=IqoPAQAAMAAJ} {\emph {\bibinfo
  {title} {Nonlinear Optics}}}\ (\bibinfo  {publisher} {Academic Press},\
  \bibinfo {year} {1992})\BibitemShut {NoStop}%
\bibitem [{\citenamefont {Bloembergen}(1996)}]{bloembergen96:book}%
  \BibitemOpen
  \bibfield  {author} {\bibinfo {author} {\bibfnamefont {N.}~\bibnamefont
  {Bloembergen}},\ }\href {https://books.google.de/books?id=VdwBT13l05EC}
  {\emph {\bibinfo {title} {Nonlinear Optics}}}\ (\bibinfo  {publisher} {World
  Scientific},\ \bibinfo {year} {1996})\BibitemShut {NoStop}%
\bibitem [{\citenamefont {Br{\'e}e}(2012)}]{bree12:book}%
  \BibitemOpen
  \bibfield  {author} {\bibinfo {author} {\bibfnamefont {C.}~\bibnamefont
  {Br{\'e}e}},\ }\href@noop {} {\emph {\bibinfo {title} {Nonlinear optics in
  the filamentation regime}}}\ (\bibinfo  {publisher} {Springer Science \&
Business Media},\ \bibinfo {year} {2012})\BibitemShut {NoStop}%
\bibitem [{\citenamefont {Hofmann}\ \emph {et~al.}(2015)\citenamefont
  {Hofmann}, \citenamefont {Hyyti}, \citenamefont {Birkholz}, \citenamefont
  {Bock}, \citenamefont {Das}, \citenamefont {Grunwald}, \citenamefont
  {Hoffmann}, \citenamefont {Nagy}, \citenamefont {Demircan}, \citenamefont
  {Jup\'{e}}, \citenamefont {Ristau}, \citenamefont {Morgner}, \citenamefont
  {Br\'{e}e}, \citenamefont {Woerner}, \citenamefont {Elsaesser},\ and\
  \citenamefont {Steinmeyer}}]{hofmann15}%
  \BibitemOpen
  \bibfield  {author} {\bibinfo {author} {\bibfnamefont {M.}~\bibnamefont
  {Hofmann}}, \bibinfo {author} {\bibfnamefont {J.}~\bibnamefont {Hyyti}},
  \bibinfo {author} {\bibfnamefont {S.}~\bibnamefont {Birkholz}}, \bibinfo
  {author} {\bibfnamefont {M.}~\bibnamefont {Bock}}, \bibinfo {author}
  {\bibfnamefont {S.~K.}\ \bibnamefont {Das}}, \bibinfo {author} {\bibfnamefont
  {R.}~\bibnamefont {Grunwald}}, \bibinfo {author} {\bibfnamefont
  {M.}~\bibnamefont {Hoffmann}}, \bibinfo {author} {\bibfnamefont
  {T.}~\bibnamefont {Nagy}}, \bibinfo {author} {\bibfnamefont {A.}~\bibnamefont
  {Demircan}}, \bibinfo {author} {\bibfnamefont {M.}~\bibnamefont {Jup\'{e}}},
  \bibinfo {author} {\bibfnamefont {D.}~\bibnamefont {Ristau}}, \bibinfo
  {author} {\bibfnamefont {U.}~\bibnamefont {Morgner}}, \bibinfo {author}
  {\bibfnamefont {C.}~\bibnamefont {Br\'{e}e}}, \bibinfo {author}
  {\bibfnamefont {M.}~\bibnamefont {Woerner}}, \bibinfo {author} {\bibfnamefont
  {T.}~\bibnamefont {Elsaesser}},\ and\ \bibinfo {author} {\bibfnamefont
  {G.}~\bibnamefont {Steinmeyer}},\ }\bibfield  {title} {\bibinfo {title}
  {Noninstantaneous polarization dynamics in dielectric media},\ }\href
  {https://doi.org/10.1364/OPTICA.2.000151} {\bibfield  {journal} {\bibinfo
  {journal} {Optica}\ }\textbf {\bibinfo {volume} {2}},\ \bibinfo {pages} {151}
(\bibinfo {year} {2015})}\BibitemShut {NoStop}%
\bibitem [{\citenamefont {Sommer}\ \emph {et~al.}(2016)\citenamefont {Sommer},
  \citenamefont {Bothschafter}, \citenamefont {Sato}, \citenamefont {Jakubeit},
  \citenamefont {Latka}, \citenamefont {Razskazovskaya}, \citenamefont
  {Fattahi}, \citenamefont {Jobst}, \citenamefont {Schweinberger},
  \citenamefont {Shirvanyan} \emph {et~al.}}]{sommer16}%
  \BibitemOpen
  \bibfield  {author} {\bibinfo {author} {\bibfnamefont {A.}~\bibnamefont
  {Sommer}}, \bibinfo {author} {\bibfnamefont {E.}~\bibnamefont
  {Bothschafter}}, \bibinfo {author} {\bibfnamefont {S.}~\bibnamefont {Sato}},
  \bibinfo {author} {\bibfnamefont {C.}~\bibnamefont {Jakubeit}}, \bibinfo
  {author} {\bibfnamefont {T.}~\bibnamefont {Latka}}, \bibinfo {author}
  {\bibfnamefont {O.}~\bibnamefont {Razskazovskaya}}, \bibinfo {author}
  {\bibfnamefont {H.}~\bibnamefont {Fattahi}}, \bibinfo {author} {\bibfnamefont
  {M.}~\bibnamefont {Jobst}}, \bibinfo {author} {\bibfnamefont
  {W.}~\bibnamefont {Schweinberger}}, \bibinfo {author} {\bibfnamefont
  {V.}~\bibnamefont {Shirvanyan}}, \emph {et~al.},\ }\bibfield  {title}
  {\bibinfo {title} {Attosecond nonlinear polarization and light--matter energy
  transfer in solids},\ }\href@noop {} {\bibfield  {journal} {\bibinfo
  {journal} {Nature}\ }\textbf {\bibinfo {volume} {534}},\ \bibinfo {pages}
{86} (\bibinfo {year} {2016})}\BibitemShut {NoStop}%
\bibitem [{\citenamefont {Quesada}\ \emph {et~al.}(2020)\citenamefont
  {Quesada}, \citenamefont {Triginer}, \citenamefont {Vidrighin},\ and\
  \citenamefont {Sipe}}]{quesada20-quant-prop-eqs}%
  \BibitemOpen
  \bibfield  {author} {\bibinfo {author} {\bibfnamefont {N.}~\bibnamefont
  {Quesada}}, \bibinfo {author} {\bibfnamefont {G.}~\bibnamefont {Triginer}},
  \bibinfo {author} {\bibfnamefont {M.~D.}\ \bibnamefont {Vidrighin}},\ and\
  \bibinfo {author} {\bibfnamefont {J.~E.}\ \bibnamefont {Sipe}},\ }\bibfield
  {title} {\bibinfo {title} {Theory of high-gain twin-beam generation in
  waveguides: From {Maxwell's} equations to efficient simulation},\ }\href
  {https://doi.org/10.1103/PhysRevA.102.033519} {\bibfield  {journal} {\bibinfo
   {journal} {Phys. Rev. A}\ }\textbf {\bibinfo {volume} {102}},\ \bibinfo
  {pages} {033519} (\bibinfo {year} {2020})}\BibitemShut {NoStop}%
\bibitem [{\citenamefont {Drummond}\ and\ \citenamefont
  {Hillery}(2014)}]{drummond14:book}%
  \BibitemOpen
  \bibfield  {author} {\bibinfo {author} {\bibfnamefont {P.}~\bibnamefont
  {Drummond}}\ and\ \bibinfo {author} {\bibfnamefont {M.}~\bibnamefont
  {Hillery}},\ }\href {https://books.google.de/books?id=YEsHAwAAQBAJ} {\emph
  {\bibinfo {title} {The Quantum Theory of Nonlinear Optics}}}\ (\bibinfo
  {publisher} {Cambridge University Press},\ \bibinfo {year}
  {2014})\BibitemShut {NoStop}%
\bibitem [{\citenamefont {Agrawal}(2007)}]{agrawal:book}%
  \BibitemOpen
  \bibfield  {author} {\bibinfo {author} {\bibfnamefont {G.P.}\ \bibnamefont
  {Agrawal}},\ }\href@noop {} {\emph {\bibinfo {title} {Nonlinear Fiber
  Optics}}},\ \bibinfo {edition} {5th}\ ed.\ (\bibinfo  {publisher}
  {Elsevier},\ \bibinfo {year} {2007})\BibitemShut {NoStop}%
\bibitem [{\citenamefont {Leuthold}\ \emph {et~al.}(2010)\citenamefont
  {Leuthold}, \citenamefont {Koos},\ and\ \citenamefont {Freude}}]{leuthold10}%
  \BibitemOpen
  \bibfield  {author} {\bibinfo {author} {\bibfnamefont {J.}~\bibnamefont
  {Leuthold}}, \bibinfo {author} {\bibfnamefont {C.}~\bibnamefont {Koos}}, \
  and\ \bibinfo {author} {\bibfnamefont {W.}~\bibnamefont {Freude}},\
  }\bibfield  {title} {\enquote {\bibinfo {title} {Nonlinear silicon
  photonics},}\ }\href {\doibase 10.1038/nphoton.2010.185} {\bibfield
  {journal} {\bibinfo  {journal} {Nature Photonics}\ }\textbf {\bibinfo
  {volume} {4}},\ \bibinfo {pages} {535--544} (\bibinfo {year}
  {2010})}\BibitemShut {NoStop}%
\bibitem [{\citenamefont {Motojima}\ \emph {et~al.}(2019)\citenamefont
  {Motojima}, \citenamefont {Suzuki}, \citenamefont {Shigekawa}, \citenamefont
  {Kainuma}, \citenamefont {An},\ and\ \citenamefont {Hase}}]{motojima19}%
  \BibitemOpen
  \bibfield  {author} {\bibinfo {author} {\bibfnamefont {Mari}\ \bibnamefont
  {Motojima}}, \bibinfo {author} {\bibfnamefont {Takara}\ \bibnamefont
  {Suzuki}}, \bibinfo {author} {\bibfnamefont {Hidemi}\ \bibnamefont
  {Shigekawa}}, \bibinfo {author} {\bibfnamefont {Yuta}\ \bibnamefont
  {Kainuma}}, \bibinfo {author} {\bibfnamefont {Toshu}\ \bibnamefont {An}}, \
  and\ \bibinfo {author} {\bibfnamefont {Muneaki}\ \bibnamefont {Hase}},\
  }\bibfield  {title} {\enquote {\bibinfo {title} {Giant nonlinear optical
  effects induced by nitrogen-vacancy centers in diamond crystals},}\
  }\href@noop {} {\bibfield  {journal} {\bibinfo  {journal} {Opt. Express}\
  }\textbf {\bibinfo {volume} {27}},\ \bibinfo {pages} {32217--32227} (\bibinfo
  {year} {2019})}\BibitemShut {NoStop}%
\bibitem [{\citenamefont {Troj{\'a}nek}\ \emph {et~al.}(2010)\citenamefont
  {Troj{\'a}nek}, \citenamefont {{\v{Z}}{\'\i}dek}, \citenamefont
  {Dzur{\v{n}}{\'a}k}, \citenamefont {Koz{\'a}k},\ and\ \citenamefont
  {Mal{\`y}}}]{trojanek10}%
  \BibitemOpen
  \bibfield  {author} {\bibinfo {author} {\bibfnamefont {F}~\bibnamefont
  {Troj{\'a}nek}}, \bibinfo {author} {\bibfnamefont {K}~\bibnamefont
  {{\v{Z}}{\'\i}dek}}, \bibinfo {author} {\bibfnamefont {B}~\bibnamefont
  {Dzur{\v{n}}{\'a}k}}, \bibinfo {author} {\bibfnamefont {M}~\bibnamefont
  {Koz{\'a}k}}, \ and\ \bibinfo {author} {\bibfnamefont {P}~\bibnamefont
  {Mal{\`y}}},\ }\bibfield  {title} {\enquote {\bibinfo {title} {Nonlinear
  optical properties of nanocrystalline diamond},}\ }\href@noop {} {\bibfield
  {journal} {\bibinfo  {journal} {Opt. Express}\ }\textbf {\bibinfo {volume}
  {18}},\ \bibinfo {pages} {1349--1357} (\bibinfo {year} {2010})}\BibitemShut
  {NoStop}%
\bibitem [{\citenamefont {Michinobu}\ \emph {et~al.}(2005)\citenamefont
  {Michinobu}, \citenamefont {May}, \citenamefont {Lim}, \citenamefont
  {Boudon}, \citenamefont {Gisselbrecht}, \citenamefont {Seiler}, \citenamefont
  {Gross}, \citenamefont {Biaggio},\ and\ \citenamefont
  {Diederich}}]{michinobu05}%
  \BibitemOpen
  \bibfield  {author} {\bibinfo {author} {\bibfnamefont {Tsuyoshi}\
  \bibnamefont {Michinobu}}, \bibinfo {author} {\bibfnamefont {Joshua~C}\
  \bibnamefont {May}}, \bibinfo {author} {\bibfnamefont {Jin~H}\ \bibnamefont
  {Lim}}, \bibinfo {author} {\bibfnamefont {Corinne}\ \bibnamefont {Boudon}},
  \bibinfo {author} {\bibfnamefont {Jean-Paul}\ \bibnamefont {Gisselbrecht}},
  \bibinfo {author} {\bibfnamefont {Paul}\ \bibnamefont {Seiler}}, \bibinfo
  {author} {\bibfnamefont {Maurice}\ \bibnamefont {Gross}}, \bibinfo {author}
  {\bibfnamefont {Ivan}\ \bibnamefont {Biaggio}}, \ and\ \bibinfo {author}
  {\bibfnamefont {Fran{\c{c}}ois}\ \bibnamefont {Diederich}},\ }\bibfield
  {title} {\enquote {\bibinfo {title} {A new class of organic donor--acceptor
  molecules with large third-order optical nonlinearities},}\ }\href@noop {}
  {\bibfield  {journal} {\bibinfo  {journal} {Chemical communications}\ ,\
  \bibinfo {pages} {737--739}} (\bibinfo {year} {2005})}\BibitemShut {NoStop}%
\bibitem [{\citenamefont {Esembeson}\ \emph {et~al.}(2008)\citenamefont
  {Esembeson}, \citenamefont {Scimeca}, \citenamefont {Michinobu},
  \citenamefont {Diederich},\ and\ \citenamefont {Biaggio}}]{esembeson08}%
  \BibitemOpen
  \bibfield  {author} {\bibinfo {author} {\bibfnamefont {Bweh}\ \bibnamefont
  {Esembeson}}, \bibinfo {author} {\bibfnamefont {Michelle~L}\ \bibnamefont
  {Scimeca}}, \bibinfo {author} {\bibfnamefont {Tsuyoshi}\ \bibnamefont
  {Michinobu}}, \bibinfo {author} {\bibfnamefont {Fran{\c{c}}ois}\ \bibnamefont
  {Diederich}}, \ and\ \bibinfo {author} {\bibfnamefont {Ivan}\ \bibnamefont
  {Biaggio}},\ }\bibfield  {title} {\enquote {\bibinfo {title} {A high-optical
  quality supramolecular assembly for third-order integrated nonlinear
  optics},}\ }\href@noop {} {\bibfield  {journal} {\bibinfo  {journal}
  {Advanced materials}\ }\textbf {\bibinfo {volume} {20}},\ \bibinfo {pages}
  {4584--4587} (\bibinfo {year} {2008})}\BibitemShut {NoStop}%
\bibitem [{\citenamefont {Nielsen}\ and\ \citenamefont
  {Chuang}(2010)}]{nielsen:book}%
  \BibitemOpen
  \bibfield  {author} {\bibinfo {author} {\bibfnamefont {M.}~\bibnamefont
  {Nielsen}}\ and\ \bibinfo {author} {\bibfnamefont {I.}~\bibnamefont
  {Chuang}},\ }\href {https://books.google.de/books?id=-s4DEy7o-a0C} {\emph
  {\bibinfo {title} {Quantum Computation and Quantum Information: 10th
  Anniversary Edition}}}\ (\bibinfo  {publisher} {Cambridge University Press},\
  \bibinfo {year} {2010})\BibitemShut {NoStop}%
\bibitem [{\citenamefont {Mandel}\ \emph {et~al.}(1995)\citenamefont {Mandel},
  \citenamefont {Wolf},\ and\ \citenamefont {Press}}]{mandel:book}%
  \BibitemOpen
  \bibfield  {author} {\bibinfo {author} {\bibfnamefont {L.}~\bibnamefont
  {Mandel}}, \bibinfo {author} {\bibfnamefont {E.}~\bibnamefont {Wolf}},\ and\
  \bibinfo {author} {\bibfnamefont {C.~U.}\ \bibnamefont {Press}},\ }\href
  {https://books.google.de/books?id=FeBix14iM70C} {\emph {\bibinfo {title}
  {Optical Coherence and Quantum Optics}}},\ EBL-Schweitzer\ (\bibinfo
  {publisher} {Cambridge University Press},\ \bibinfo {year}
  {1995})\BibitemShut {NoStop}%
\end{thebibliography}
%

\end{document}